# Experimental Observation of High Intrinsic Thermal Conductivity of AlN

Zhe Cheng,[1] Yee Rui Koh,[2] Abdullah Mamun,[3] Jingjing Shi,[1] Tingyu Bai,[4] Kenny Huynh,[4] Luke Yates,[1] Zeyu Liu,[5] Ruiyang Li,[5] Eungkyu Lee,[5] Michael Liao,[4] Yekan Wang,[4] Hsuan Ming Yu,[4] Maki Kushimoto,[6] Tengfei Luo,[5] Mark S. Goorsky,[4] Patrick E. Hopkins,[2,8,9] Hiroshi Amano,[7] Asif Khan,[3] Samuel Graham[1,10,*]

[1] George W. Woodruff School of Mechanical Engineering, Georgia Institute of Technology, Atlanta, Georgia 30332, United States

[2] Department of Mechanical and Aerospace Engineering, University of Virginia, Charlottesville, Virginia 22904, United States

[3] Department of Electrical Engineering, University of South Carolina, Columbia, South Carolina 29208, United States

[4] Materials Science and Engineering, University of California, Los Angeles, Los Angeles, CA, 91355, United States

[5] Department of Aerospace and Mechanical Engineering, University of Notre Dame, Notre Dame, Indiana 46556, United States

[6] Department of Electrical Engineering and Computer Science, Nagoya University, Furo-cho, Chikusa-ku 464-8603, Nagoya, Japan

[7] Institute of Materials and Systems for Sustainability, Nagoya University, Furo-cho, Chikusa-ku 464-8601, Nagoya, Japan

[8] Department of Materials Science and Engineering, University of Virginia, Charlottesville, Virginia 22904, United States

[9] Department of Physics, University of Virginia, Charlottesville, Virginia 22904, United States

[10] School of Materials Science and Engineering, Georgia Institute of Technology, Atlanta, Georgia 30332, United States

[*] Corresponding author: sgraham@gatech.edu

**Abstract**

AlN is an ultra-wide bandgap semiconductor which has been developed for applications including power electronics and optoelectronics. Thermal management of these applications is the key for stable device performance and allowing for long lifetimes. AlN, with its potentially high thermal conductivity, can play an important role serving as a dielectric layer, growth substrate, and heat spreader to improve device performance. However, the intrinsic high thermal conductivity of bulk AlN predicted by theoretical calculations has not been experimentally observed because of the difficulty in producing materials with low vacancy and impurity levels, and other associated defect complexes in AlN which can decrease the thermal conductivity. This work reports the growth of thick (>15 μm) AlN layers by metal-organic chemical vapor deposition with an air-pocketed AlN layer and the first experimental observation of intrinsic thermal conductivity from 130 K to 480 K that matches density-function-theory calculations for single crystal AlN, producing some of the highest values ever measured. Detailed material characterizations confirm the high quality of these AlN samples with one or two orders of magnitude lower impurity concentrations than seen in commercially available bulk AlN. Measurements of these commercially available bulk AlN substrates from 80 K to 480 K demonstrated a lower thermal conductivity, as expected. A theoretical thermal model is built to interpret the measured temperature dependent thermal conductivity. Our work demonstrates that it is possible to obtain theoretically high values of thermal conductivity in AlN and such films may impact the thermal management and reliability of future electronic and optoelectronics devices.

## Introduction

Aluminum nitride (AlN) is an ultra-wide bandgap semiconductor which has been developed for applications of power electronics and optoelectronics such as next-generation ultraviolet (UV) light-emitting diodes (LED) and laser diodes.[1] The thermal management of these devices is the key for stable performance and long lifetime, especially for high-power and high-frequency power electronics and high-power optoelectronics. For instance, the external quantum efficiency of deep UVC LEDs is typically in the single-digit percent range, even less than 1% for LEDs with wavelength shorter than 250 nm.[2,3] The majority of the input power is converted to Joule-heating which raises the LED temperature and forms the hotspot in the junction area. The device temperature is a critical factor which significantly affects the wavelength, reliability, and lifetime and correspondingly limits the maximum output power of UV LEDs.[2] Moreover, wide bandgap materials as substrates are critical for UVC LEDs in order to provide the necessary optical transparency for light extraction. In other applications beyond optoelectronics, wide bandgap materials with hexagonal crystal structures are used as interfacial layers or growth substrates in devices requiring the heteroepitaxial growth of wide bandgap power or radio frequency semiconductors like GaN and $Ga_2O_3$, both also require similar attention to manage their temperature rise. Therefore, materials with wide bandgaps and high thermal conductivity such as AlN are of importance in several technologically important applications which require efficient heat dissipation to ensure proper device operation.

Even though it was first synthesized more than one century ago, the thermal conductivity of AlN was first reported to be about 1.76 W/m-K at room temperature in 1959, and about 30 W/m-K at 473 K in the form of hot-pressed powder in 1960 due to the delayed development of thermal characterization techniques.[4,5] Until 1973, the thermal conductivity of single crystal AlN was reported to be 200 W/m-K at room temperature with high concentrations of oxygen impurities and later increased to 285 W/m-K with improved quality in 1987.[6-8] After that, more thermal studies were performed on AlN ceramics and commercial AlN substrates but no higher thermal conductivity values were observed.[9-13] This value (285 W/m-K) has been widely used as the room-temperature thermal conductivity of bulk AlN for decades. Even though the calculated intrinsic thermal conductivity was reported by Slack in 1973 to be 320 W/m-K.[6] This higher value was obtained by extrapolating the value of thermal conductivity to the case where the oxygen impurity concentrations is zero. However, no experimental observation of this high value of thermal conductivity has been reported because of the difficulty in growing AlN materials with such low vacancy and impurity levels, and addressing other associated defect complexes in AlN.[14-18]

This work, for the first time, demonstrates a high-quality and high-purity AlN film which meets the previous theoretical predictions of high thermal conductivity. The thick single-crystal AlN films were grown by metal-organic chemical vapor deposition (MOCVD) on sapphire substrates with an air-pocketed AlN layer at the sapphire interface. This layer allows for the growth of thick AlN without any cracking. The thermal conductivity was determined using time-domain thermoreflectance (TDTR) while density function theory (DFT) was used to calculate the thermal conductivity of perfect single crystal AlN to compare with the measured results. Additionally, the thermal conductivity of two commercially-available single-crystal AlN substrates were also measured to compare with the MOCVD samples. Detailed material characterization including scanning transmission electron microscopy (STEM), x-ray diffraction (XRD), and secondary-ion mass spectrometry (SIMS) were used to verify the crystal quality and purity and help to interpret the measured thermal conductivity data. Finally, a thermal model was also built to understand the origin of the reduced thermal conductivity of the two commercially-available single-crystal AlN substrates.

**Results and Discussion**

For this study, four separate AlN samples were chosen for thermal conductivity measurements. Samp_1 and Samp_2 were grown by MOCVD on sapphire substrates with slightly different AlN thicknesses, being 18 μm and 22.5 μm, respectively. More on the MOCVD growth details can be found in the Methods Section. Samp_3 and Samp_4 (~500 μm thick wafers) are purchased from HexaTech, Inc. and grown by physical vapor transport (PVT).[19] As shown in Figure 1(a), the AlN studied in this work has a wurtzite crystal structure and four atoms in one unit cell. Figure 1(b) shows the Raman spectrum of Samp_2. The Raman peaks are A1 (TO) with phonon frequency of 18.39 THz, E2 with phonon frequency of 19.73 THz, E1 (TO) with phonon frequency of 20.09 THz, and A1 (LO) with phonon frequency of 26.66 THz.[20] The other three peaks are from the sapphire substrates.[21] All the Raman peaks are clean and sharp. This observation verified that the AlN film is high quality, and no intermixing of the impurities and substances present in the film. Figure 1(c) shows the phonon dispersion relation of AlN calculated by DFT. AlN has three acoustic phonon branches and nine optical branches. The phonon frequencies of polarized materials approaching the gamma point from different directions are different which result in the frequency discontinuous points at the gamma point. In terms of Al$X$ compounds ($X$ = N, P, As, and Sb), AlN has the largest acoustic phonon frequency scale and the highest acoustic velocities, resulting in the high thermal conductivity.[22] There is no appreciable phonon band gap between acoustic and optical phonon branches of AlN which enables extensive acoustic phonon- optical phonon scattering which decreases the thermal conductivity of AlN rapidly with increasing temperature at relatively high temperatures.[22]

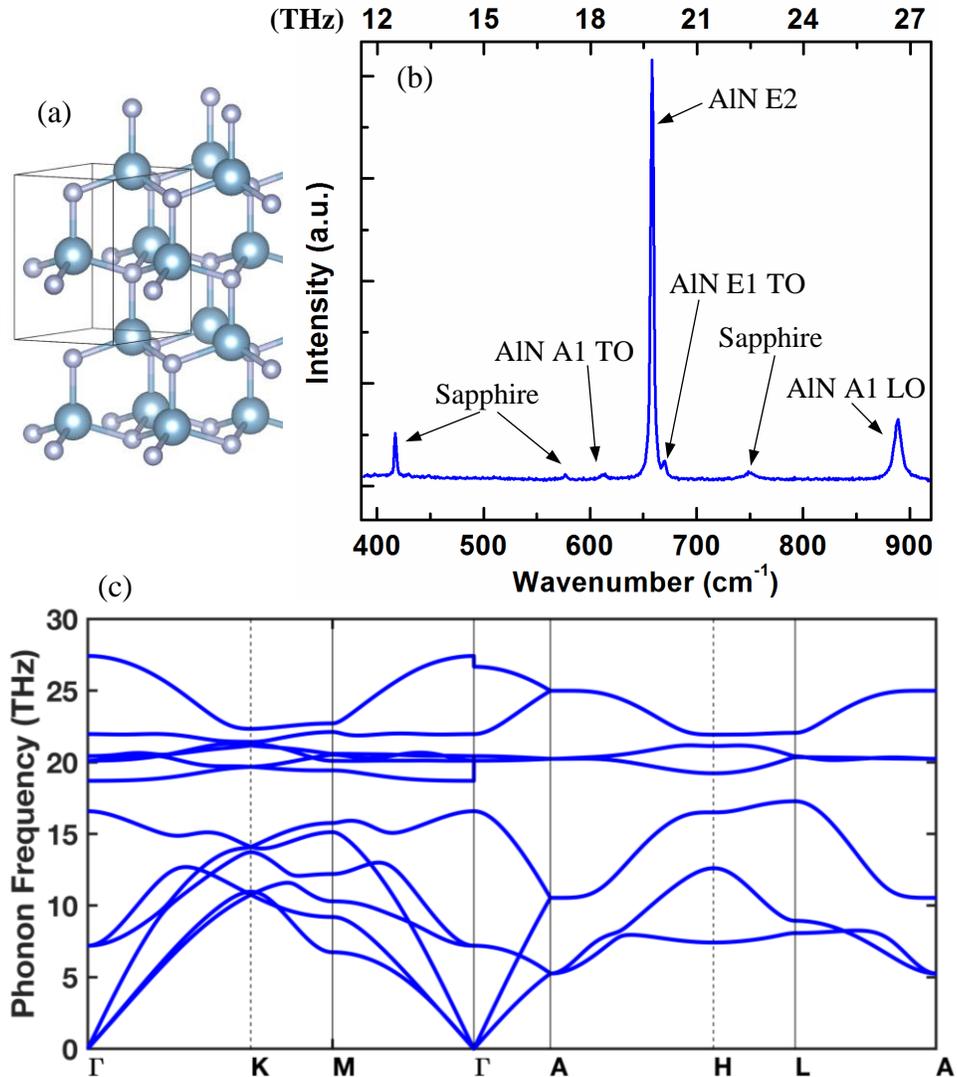

Figure 1. (a) crystal structure of AlN. (b) Raman spectroscopy of Samp_2. (c) phonon dispersion relation of AlN calculated by DFT.

Figure 2(a) shows the high-resolution STEM image along (1$\bar{1}$00) zone axis of Samp_1, depicting high-quality AlN and clear lattices. To study the structure of the MOCVD grown AlN, Figure 2(b) shows the near-surface cross-sectional STEM image of AlN of Samp_1. A few dislocations show up along the growth direction. Figure 2(c) shows the near-surface plan-view STEM image of Samp_1 with some dislocations. The dislocation density of Samp_1 and Samp_2 which were grown on sapphire is estimated to be 1.6 x $10^8$/cm$^2$ while that of Samp_3 and Samp_4 (bulk materials) are several orders of magnitude lower. More details can be found in Supplemental information (Figure S1 and S2). While the dislocation density is higher in the hetero-epitaxially MOCVD grown material than the bulk PVT samples, we expect this dislocation density to have a negligible impact on the thermal conductivity.[23] Moreover, the dislocations are aligned along the C-axis direction which affects cross-plane thermal conductivity less than the in-plane thermal conductivity. Based on our laser spot size, our TDTR measurements are more sensitive to the cross-plane thermal conductivity. To compare the impurity concentrations in these samples, SIMS is used to measure the concentrations of carbon (C), oxygen (O), and silicon (Si) of Samp_1, Samp_3, and Samp_4. As shown in Figure 2(d), the impurity concentrations of Samp_1 is one or two orders of magnitude lower than that of Samp_3 and Samp_4, indicating the higher purity of MOCVD samples.

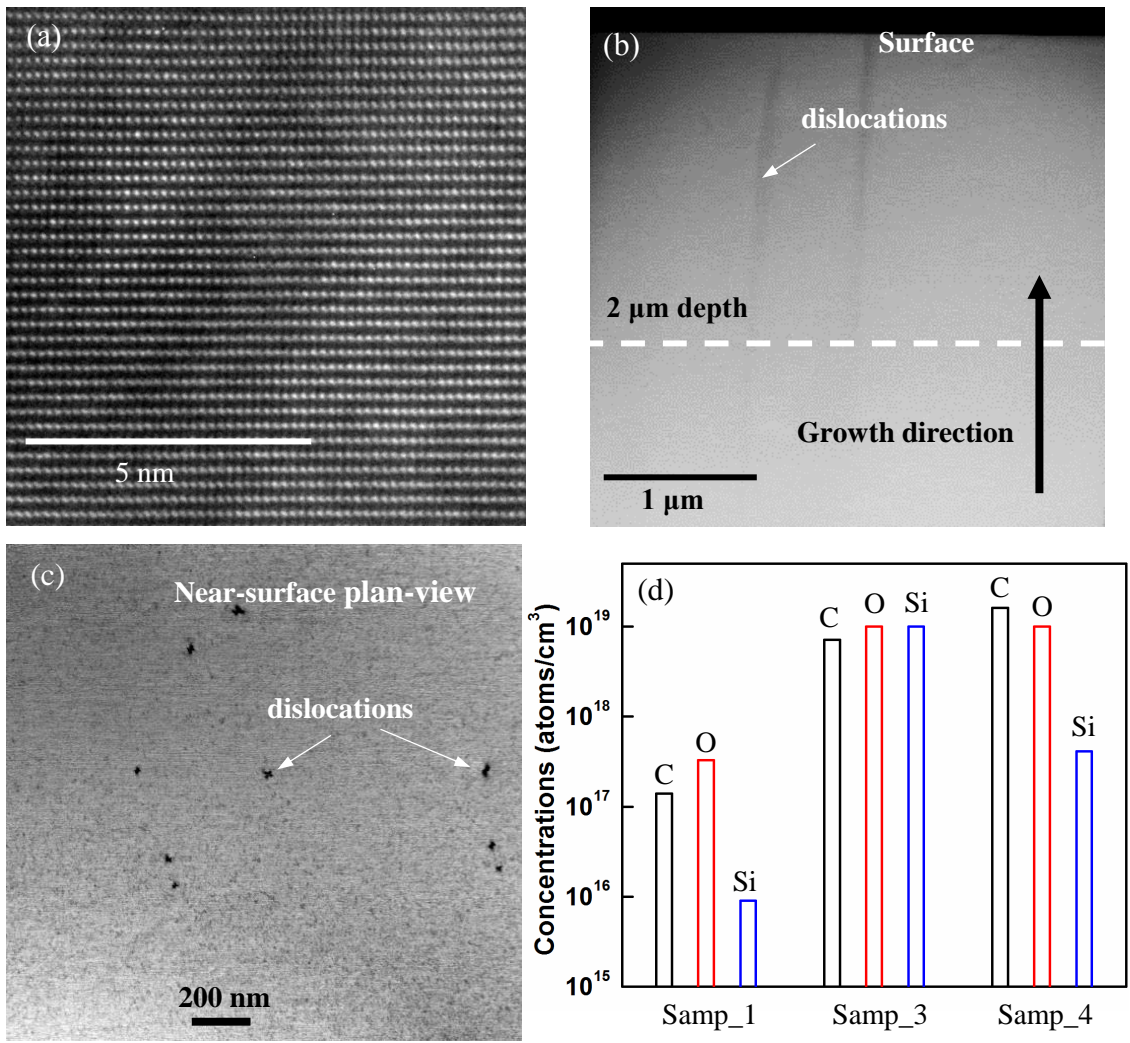

Figure 2. (a) HRSTEM image along (1$\bar{1}$00) zone axis of Samp_1. (b) Near-surface cross-section image of AlN grown on sapphire substrates of Samp_1. (c) Near-surface plan-view STEM image of Samp_1. (d) Impurity concentrations of carbon (C), oxygen (O), and silicon (Si) of Samp_1, Samp_3, and Samp_4 by SIMS.

The temperature dependent thermal conductivity of Samp_1 and Samp_2 are shown in Figure 3(a) and compared with DFT-calculated thermal conductivity. The details of TDTR measurements can be found in the Methods Section. Round-robin measurements were performed by TDTR in three different laboratories (Georgia Tech, University of Virginia, and University of Notre Dame) and consistent results are obtained. The error bars are calculated with a Monte Carlo method by considering all possible errors (see Supplementary Information). The DFT calculations for the thermal conductivity at low temperatures ("DFT_Lindsay") is from the literature[22] and the DFT calculated thermal conductivity at high temperatures is from this work. Excellent agreement between experimentally measured thermal conductivity and DFT-calculated thermal conductivity is achieved from 130 K to 480 K, which highlights the high quality of the MOCVD-grown AlN and confirms that the dislocations and small amounts of impurities have a negligible impact on measured thermal conductivity in Samp_1 and Samp_2. Here, TDTR measurements only probe the top portion of the thick films and any defects near the growth interface are eliminated from the analysis even at temperatures as low as 130 K.

Figure 3(b) shows the historical reporting of experimentally measured thermal conductivity of AlN.[4,6-13] Please note that the first reported value of AlN thermal conductivity (1.76 W/m-K) in 1959 is very low so we do not plot it in Figure 3(b) considering the scale of the graph. The thermal conductivity data reported in 1960 is at 473 K in form of hot-pressed AlN powder.[4] All the other values are room-temperature values. The red line is DFT-calculated theoretical value of AlN thermal conductivity. As shown in Figure 3(b), the thermal conductivity reported in this work for Samp_1 and Samp_2 is the highest reported thermal conductivity of AlN which matches the DFT-calculated theoretical predictions. This is due to the high quality of these MOCVD grown samples (small concentrations of impurities and vacancies) which will be discussed in detail later. Our measured thermal conductivity (321 W/m-K) in this clearly shows that the widely used value of 285 W/m-K for bulk AlN can be exceeded.

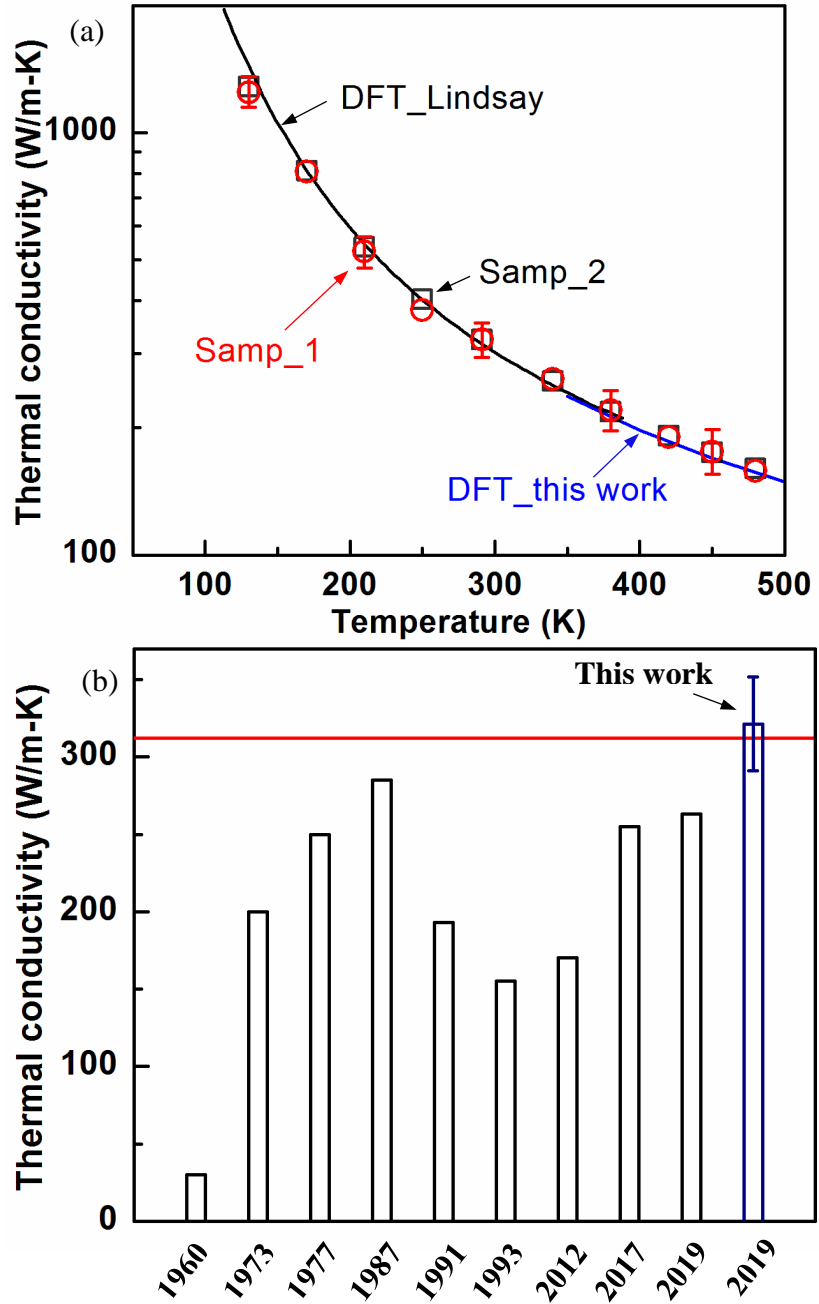

Figure 3. (a) temperature dependent thermal conductivity of Samp_1 and Samp_2, and first-principle calculated temperature dependent thermal conductivity of single crystal AlN. (b) the experimentally measured thermal conductivity of AlN reported with time.[4,6-13] The data in 1960 is at 473 K and all the other values are room-temperature values.[4] The red line is DFT-calculated theoretical value of AlN thermal conductivity.

For comparison to the two MOCVD-grown AlN samples, two commercially available AlN substrates grown by PVT were measured by TDTR (Samp_3 and Samp_4). The measured thermal conductivity of these four samples are summarized in Figure 4(a). The thermal conductivity of Samp_3 and Samp_4 are lower than the MOCVD samples at all temperature points. The thermal conductivity of Samp_3 and Samp_4 at room temperature are 278 W/m-K and 216 W/m-K, close to literature values for samples from this source.[13] It has been reported that Al vacancies play a dominant role in limiting the thermal conductivity of these commercially available PVT AlN substrates.[13] To understand the measured thermal conductivity of Samp_3 and Samp_4, we used a Callaway thermal model employing full phonon dispersion relations and Matthiessen's rule to account for different phonon scattering mechanisms. More details can be found in the Methods Section. Defect-phonon scattering rate is proportional to the square of mass difference ratio. The mass of vacancy is zero so phonon-vacancy scattering has very large mass difference ratio and corresponding phonon-vacancy scattering rate, leading to significantly reduced thermal conductivity. Our thermal model includes the effect of impurity-phonon scattering (C, O, Si) and phonon-vacancy scattering. Because it is very challenging to measure the vacancy concentration in AlN, we infer the Al vacancy concentration by fitting the temperature dependent thermal conductivity data. The Al vacancy concentrations in Samp_1 and Samp_2 are negligible. The Al vacancy concentrations for Samp_3 and Samp_4 are estimated to be $3 \times 10^{19}$ cm$^{-3}$ and $1.5 \times 10^{20}$ cm$^{-3}$, which are comparable or higher than the impurity concentrations. Please note that this is a first-order estimation because an accurate calculation method including impurities or even defect complexes are under development. Here, we assume all the impurities and vacancies are point defects and the scattering rates calculation are based on equations (3-5). We can see the large deviations between experimental data and thermal model data at low temperatures with structural imperfection scattering dominating phonon transport. We attribute the deviations at low temperatures to the limitation of the thermal model and possible other scattering mechanisms.[24] Figure 4(b) shows the scattering rates of different samples at room temperature. The defects scattering rates of all samples are lower than the phonon-phonon Umklapp scattering rates at the whole acoustic phonon frequency range. All these scattering rates increase with frequency because low frequency phonons are less scattered by other phonons and structural imperfections. The defects scattering rates of Samp_3 and Samp_4 are much higher than that of Samp_2 due to the different impurity and vacancy concentrations. Overall, combing the thermal characterizations and materials structural characterizations, Samp_1 and Samp_2 grown by MOCVD have much lower structural defect levels including impurities and vacancies than Samp_3 and Samp_4 grown by PVT.

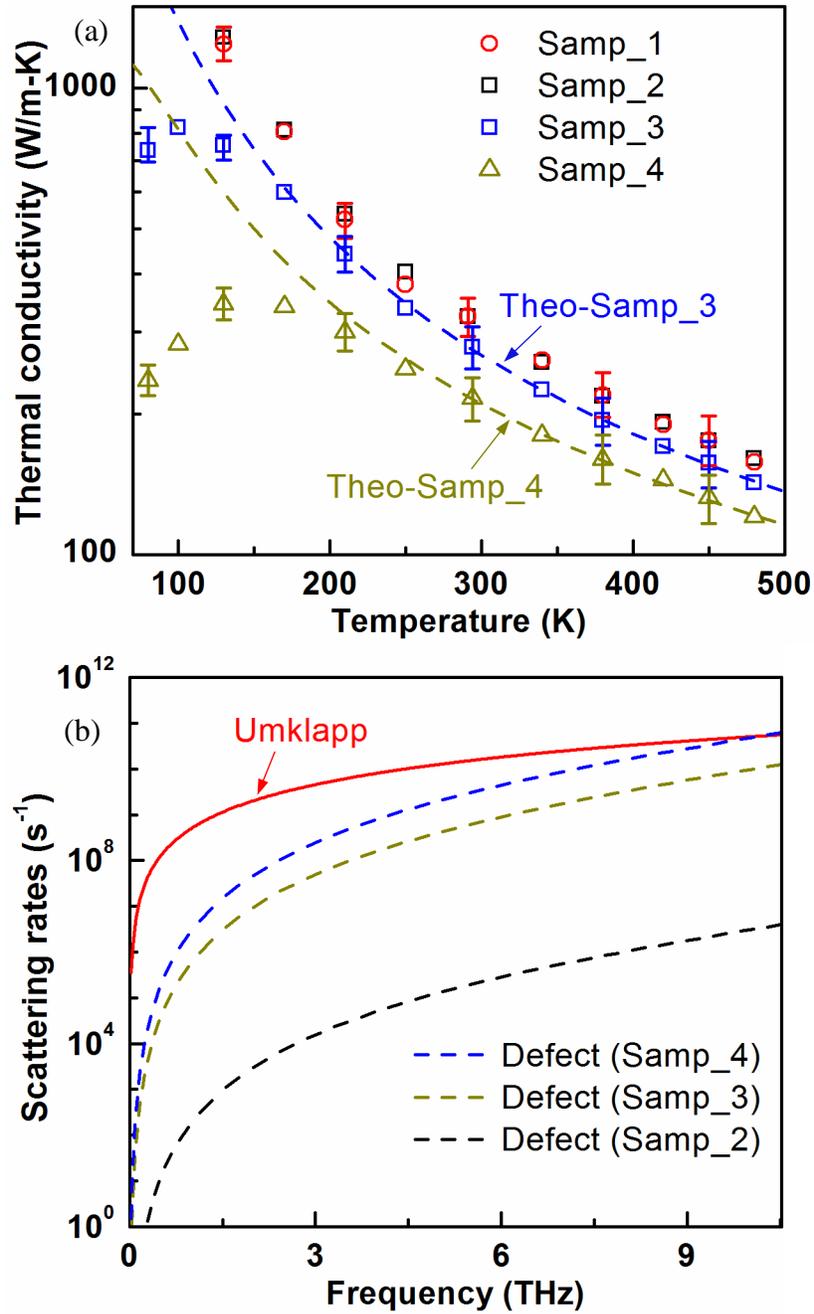

Figure 4. (a) measured temperature dependent thermal conductivity of Samp_1, Samp_2, Samp_3, and Samp_4. The dash lines are theoretical fitting curves by considering the effects of impurities and vacancies. (b) Scattering rates for different scattering sources: phonon-phonon Umklapp scattering, phonon-defects scatterings including impurities and vacancies.

## Conclusions

In this work, we reported the first experimental observation of the intrinsic thermal conductivity of AlN grown by MOCVD from 130 K to 480 K which matches excellently with DFT-predicted thermal conductivity of single crystal AlN. Detailed material characterizations show that the MOCVD AlN samples have one or two orders of magnitude lower impurity concentrations than the commercially available PVT substrates even though they have a higher dislocation density. As comparison, two commercially available PVT AlN substrates were measured from 80 K to 480 K and the measured thermal conductivity were close to literature values of samples from the same supplier. Moreover, the results of this work demonstrate that it is possible to grow thick films of high thermal conductivity AlN for heat dissipation applications in wide bandgap electronics with thermal conductivity values that exceed bulk material available today. The lower thermal conductivity values in commercially available bulk substrates are attributed to having high concentrations of Al vacancies or vacancy-impurity complexes. Our work clearly shows that MOCVD growth can reduce the level of these impurities while growing thick films with thermal properties that match theoretical predictions for the first time. This result will open new doors for research with AlN both from fundamental thermal science and to semiconductor growth, especially in applications where thermal management is of high importance.

## Methods Section

### Sample Growth

The AlN growth for this study was carried out using a custom MOCVD reactor. The growth reactor had a fast metal-organic precursor switching manifold. Two-inch diameter $0.2^0$ off-axis basal plane sapphire substrates were used for all the growths. The growth procedure started with a pulsed high-temperature (1373 K) epitaxy sequence where the Al- and N-precursors (Trimethyl Aluminum and NH$_3$) were alternately supplied to the growth chamber. The Al-N precursor flow rate and the V/III ratio are adjusted to yield air-pocketed AlN layers at the sapphire interface. The air-pockets help managing the strain and thus allow the growth of AlN layers with thicknesses well over 15 µm without any cracking. Subsequently the growth temperature is increased to 1473 K and the flows are modified to yield smooth layers. This two-step procedure leads to a significant reduction in the number of defects due to dislocation bending/annihilation and cracking is avoided due to strain relief from the air-pockets at the interface. A chamber pressure of 40 torr was maintained during the entire two-step growth process. The high growth temperatures and the purity of the growth-precursors leads to a significant reduction in the incorporation of oxygen, carbon and other impurities in our MOCVD grown layers. The layer thicknesses were measured from a cross-section image obtained using a scanning electron microscope (SEM). The surface quality was then characterized using an Atomic Force Microscope (AFM) and the RMS value of surface roughness was measured to be 0.25 nm. The off-axis (102) X-ray linewidth was measured to be around 350 arcsec. Based on our previous correlation studies, using etch pit density counts as a measure of the defects, we estimated the defects in our studied layers to be around (1-3 x 10$^8$ cm$^{-2}$).

**Thermal Characterizations**

TDTR is a pump-probe technique which can be used to measure thermal properties of both nanostructured and bulk materials.[25,26] The AlN surface was first coated with a layer of Al (~80 nm) as transducer. The local Al thickness is determined by the picosecond acoustic technique.[27] A modulated pump beam (400 nm) heats the sample surface while a delayed probe beam (800 nm) detects the temperature variation of the sample surface via thermoreflectance. The delay time is controlled by a mechanical delay stage and the signal is picked up by a photodetector and a lock-in amplifier. The pump and probe beam sizes are 19.0 μm and 13.3 μm (diameters) and the modulation frequency is 3.6 MHz. The measured signal is fitted with an analytical heat conduction solution of the multi-layer sample structure to infer unknown parameters.[26,28] For samples in this work, the unknown parameters are Al-AlN thermal boundary conductance (TBC), AlN thermal conductivity. There are three layers for Samp_1 and Samp_2. But the AlN are thermally thick for most of the temperature points, the fitting results of the three-layer model (Al + AlN + sapphire) are very close to the two layer model (Al + AlN). The heat capacity of Al, AlN, and sapphire, the thermal conductivity of Al and sapphire are from literature.[29-31] The TDTR sensitivity and data fitting can be found in the Supplementary Information. We used a Monte Carlo method to calculate the errors of these TDTR measurements and more details is included in the Supplementary Information.

**DFT Calculation**

An iterative scheme is applied to solve the linearized phonon Boltzmann transport equation with the help of first-principles force constants. We first relax the AlN atomic structure to its optimized positions using *Quantum Espresso*.[32] Then second order force constants which provide phonon frequencies, group velocity and scattering phase space, are calculated using Density Functional Perturbation Theory (DFPT), using an $8 \times 8 \times 4$ $q$ space grid. Finite different methods implemented in thirdorder.py[33] are used to calculated third order force constants in order to calculate three-phonon scattering rates based on a $4 \times 4 \times 4$ supercell. Linearized phonon BTE is solved iteratively using *ShengBTE* in a $12 \times 12 \times 12$ Monkhorst-Pack grid.[33]

**Thermal Model**

The thermal conductivity of non-metal crystalline material can be expressed as[34-37]

$$k = \frac{1}{3}\sum_p \int_0^{\omega_{\text{cut-off}}} \hbar\omega D_\lambda \frac{df_{BE}}{dT} v_\lambda^2 \tau_{C,\lambda}, \quad (1)$$

where $\omega$ is the phonon frequency, $\hbar$ is the reduced Planck constant, $v_\lambda$ is the modal phonon group velocity of phonon mode $\lambda$, $\tau_{C,\lambda}$ is the modal combined relaxation time, $\sum_p$ is over all phonon polarizations, and $D_\lambda$ is the modal phonon density of states, and $f_{BE}$ is the Bose-Einstein distribution function. The combined relaxation time $\tau_{C,\lambda}$ of each phonon mode can be obtained from the Matthiessen's rule as[36,37]

$$\tau_{C,\lambda} = \left(\frac{1}{\tau_U} + \frac{1}{\tau_M} + \frac{1}{\tau_B}\right)^{-1}, \quad (2)$$

where $\tau_U$, $\tau_M$, and $\tau_B$ are the relaxation times of Umklapp phonon-phonon scattering, mass-difference phonon-impurity scattering, and phonon-boundary scattering, respectively. The scattering rate expressions are[13,38-40]

$$\frac{1}{\tau_U} = BT\omega^2 e^{-\frac{C}{T}}, \quad (3)$$

$$\frac{1}{\tau_M} = \frac{V\omega^4}{4\pi v^3}\sum_i x_i \left(\frac{\Delta M_i}{M}\right)^2, \quad (4)$$

$$\frac{1}{\tau_B} = v/d, \quad (5)$$

where $B$ and $C$ are fitting parameters, $V$ is the volume of the AlN primitive cell, $x_i$ is the atomic fraction of sites occupied by defect $i$, $\Delta M_i$ is the mass difference between defect and original atom, and $d$ is the thickness of the AlN sample. To calculate the scattering rates of different mechanisms, the parameters $B$ and $C$ are fitted with the first-principle-calculated AlN thermal conductivity. The impurity scattering rates of C, O, and Si are calculated based on the SIMS results. Additionally, the Al vacancies are also considered and the vacancy concentrations are obtained by fitting the analytical predictions with experimental measurements.

**Raman Spectroscopy**

We employed a Reinshaw Raman InVia Microscopy to analyze the 18 µm AlN film, with a 50x objective lens and 488 nm laser. All symmetry-allowed optical phonons in AlN film is dominant by $E_2$ mode, follows by weaker $A_1$(LO), $A_1$(TO), and $E_1$(TO) modes. Weak Raman spectra for the substrate, $Al_2O_3$ can also be observed in the analysis.

## Acknowledgements

The authors would like to acknowledge the financial support from Office of Naval Research MURI Grant No. N00014-18-1-2429.

## Competing Financial Interest

The authors claim no competing financial interest.